\RequirePackage{tikz}
\documentclass[doublecol, linenumbers]{epl2.modif} 

\usepackage{amsmath}
\usepackage{amsthm}
\usepackage{mathrsfs}
\usepackage{amssymb}
\usepackage{hyperref}
\usepackage{tikz}
\usetikzlibrary{arrows,snakes,backgrounds}

\usepackage{verbatim}

\newcommand{\ov}[1]{\overline{#1}}
\newcommand{\Tr}{{\rm Tr}}
\newcommand{\de}{\partial}

\newcommand{\Sym}{\text{sym}\Big\{1, 2, 3\Big\}}

\newcommand{\R}{\mathbb{R}^3}

\title{Functional Renormalisation Group analysis of a Tensorial Group Field Theory on $\mathbb{R}^3$ }
\shorttitle{Functional Renormalisation Group analysis of a TGFT on   $\mathbb{R}^3$} 

\author{Joseph Ben Geloun\inst{1,3} \and Riccardo Martini\inst{1,2} \and Daniele Oriti\inst{1}}
\shortauthor{Functional Renormalisation Group analysis of a TGFT on   $\mathbb{R}^3$}

\institute{                    
  \inst{1} Max Planck Institute for Gravitational Physics, Albert Einstein Institute, Am M\"uhlenberg 1, 14476, Potsdam, Germany\\
  \inst{2} Alma Mater Studiorum, Universit\`a di Bologna, via Zamboni 33,
40126, Bologna, Italy \\ 
\inst{3} International Chair in Mathematical Physics and Applications,
ICMPA-UNESCO Chair, 072BP50, Cotonou, Benin
}

\abstract{
We study a model of Tensorial Group Field Theory (TGFT) on $\mathbb{R}^3$ from the point of view of the Functional Renormalisation Group. This is the first attempt to apply a renormalisation procedure to a TGFT model defined over a non-compact group manifold. 
IR divergences (with respect to the metric on $\mathbb{R}$) coming from the non-compactness of the group are regularised via compactification, and a thermodynamic limit is then taken. 
We identify then IR and UV fixed points of the RG flow and find strong hints of a phase transition of the TGFT system from a symmetric to a broken or condensate phase in the IR.}

\begin{document}

\maketitle

\section{Introduction}
Group Field Theories \cite{daniele60,10questions, thomas, GFTrecent, GFTandLQG} (GFTs) are a particular class of quantum field theories with fields defined over a group manifold and characterised by combinatorially non-local interaction terms. This combinatorial non-locality makes the Feynman diagrams of the theory stranded diagrams dual to cellular complexes (simplicial complexes in the simplest constructions) \cite{daniele60,GFTrecent}. 
GFT's historically were born from an attempt to generalise matrix models \cite{diFrancesco} of $2d$ gravity to higher dimensions  in the form of tensor models \cite{ambjorn,gross,boulatov}. These models were soon enriched with group theoretic data in such a way that the Feynman amplitudes of specific GFT models coincide with state sum models of topological field theory \cite{boulatov,ooguri}. A connection between GFTs and Loop Quantum Gravity (LQG) 
\cite{LQG, thiemann} was immediately pointed out \cite{rovelli} at the level of quantum states, and later enforced at the level of the quantum dynamics. In fact, it was later shown \cite{mikecarlo} that GFTs provide a formal (complete) definition of spin foams models, a covariant formalism for LQG. For GFTs (and spin foam models) endowed with a discrete geometric interpretation (which requires appropriate group theoretic data and suitable choices of dynamics) there is also a direct link with simplicial quantum gravity path integrals, first manifested in the semiclassical analysis of GFT Feynman (spin foam) amplitudes, where one recovers the Regge action \cite{SFasymp}, and then shown to be generically manifest in the flux representation of the same amplitudes \cite{aristidedanieleHost}.

One key open issue of all of the above quantum gravity approaches, and GFTs in particular, is the emergence of a continuous geometry out of the discrete and quantum pre-geometric structures defining the formalisms, and of General relativity as an effective description of their (collective) dynamics in the same approximation. The study of the GFT collective dynamics and of the associated continuum limit is therefore crucial. Moreover, one suggested scenario for the emergence of spacetime and geometry out of such quantum gravity models involves a phase transition (dubbed ``geometrogenesis'' \cite{graphity}) 
from a pre-geometric phase to a geometric phase, which may be further identified to a condensate phase of the underlying quantum gravity system \cite{GFTfluid,danieleEmergence} (a similar idea was proposed also in a loop quantum gravity context in \cite{tim}). Indeed, GFT condensate states seem to possess an effective dynamics with a cosmological interpretation \cite{GFTcondensate}. Within this perspective, the advantage of the GFT formalism is that it offers the possibility to address this issue with tools coming from standard quantum field theory.  

And indeed, an important line of recent developments has concerned the renormalisation of GFT models, since the renormalisation group is indeed the key tool to address both the quantum consistency of field theory dynamics and the definition of the continuum limit, aimed at a precise mapping of the phase diagram of the theory.  Furthermore, GFT renormalization is also one of the two main strategies to define and study the renormalization of spin foam models, the other being through a generalised lattice gauge theory approach \cite{SFren}. Most work in this direction has concerned a particular class of GFTs, called Tensorial Group Field Theories (TGFT's) \cite{bgr, Car, BenGeloun:2012pu, kra, cor1, dine1, cor2, rankd, Sylvain, Joseph, danieleLahoche},
which incorporate recent advances 
in the statistical analysis of colored tensor models \cite{bgrr,bgriv,gurauryan,bgr2}. In particular, in TGFT framework, fields are endowed with tensorial transformation properties under the action of the group itself. The perturbative analysis of these field theories
has been undertaken and a large set of models prove to be perturbatively renormalisable and asymptotically free (see references above). 
However, understanding the continuum limit, including the phase diagram and phase transitions of the same models requires the study of
their non-perturbative properties. 

This being the goal, the Functional Renormalisation Group is  an efficient
framework to reach it \cite{delamotte,wetterich,wetterich2,morris}. 

The FRG approach has been applied first to matrix models \cite{Brezin:1992yc,TimAstridI,TimAstridII} 
(with the double scaling critical point re-interpreted as a fixed point in the RG flow). More recently, the FRG framework has been adapted and applied for the first time to TGFTs in \cite{dariojosephdaniele}. 
The authors of \cite{dariojosephdaniele} studied a rank-3 TGFT defined over a compact $U(1)$ group manifold. The $\beta$-functions define a non-autonomous system in the cut-off $N$. Then, the authors studied
two regimes of the cut-off, the large and small $N$ (and also an intermediate
regime at fixed $N$), where a proper
notion of dimension of the couplings can be defined and an autonomous system of RG equations is obtained. The notion of UV or IR 
``fixed points'' is then only loosely (i.e. asymptotically) defined, as the existence of a trajectory from a UV to a IR fixed point becomes
more difficult to ascertain. This is not surprising nor problematic per se, and it simply signals the presence of an additional scale in the formalism, here the size of the group manifold on which the fields are defined. In fact, the same feature is found in different contexts like quantum field theory at finite temperature, on 
non-commutative manifolds and on a curved spacetimes (see \cite{dariocurved} and references therein). Still, hints of a phase transition from a symmetric to a broken phase, in the approximation of large size of the group manifold, were found.

Note that progress towards the a better characterisation of the phase diagram and of phase transitions in tensor models has also been recently achieved \cite{Delepouve:2015nia,Benedetti:2015ara}. In fact, using a similar mode integration alongside double scaling limit techniques, the nonperturbative analysis of quartic tensor models has been performed, with evidence of spontaneous symmetry breaking mechanism similar to that suggested in \cite{dariojosephdaniele}. The models considered (quartic tensor models with trivial kinetic term) as well as the techniques employed (double scaling and intermediate field representation, allowing to solve quartic tensor models with matrix models methods) are very different from the ones employed in our present study and in \cite{dariojosephdaniele}, which are based on generic FRG conventions and concern TGFT models with non trivial kinetic kernels.  

In this work, we study a class of TGFT models which possess no such additional scale, thus are expected to show proper fixed points, so, in a sense, improve on the previous analysis. The model we consider is a rank-3 TGFT with fields defined on the non-compact manifold $\R$, and endowed with a Laplacian kinetic term. Being a TGFT, this model is of interest as a toy model for quantum gravity, due to the combinatorics of its Feynman diagrams, but even more because it can be seen as a (much) simplified version of Lorentzian (T)GFTs for 4d quantum gravity, also based on a non-compact group manifold. Thus it can be seen as a useful exercise on the way to a renormalisation analysis of more
realistic models, hopefully providing useful hints of what to expect for them.  

One certainly generic feature is that the non-compact manifold introduces IR divergences, that we properly 
address through a careful definition of a thermodynamic limit for TGFTs, in this FRG context. This is an important technical lesson for later developments. In this limit, we recover an autonomous system of $\beta$-functions of the coupling constants, and we can then identify the UV and IR fixed points of the RG flow. We also find evidence for a phase transition (in the continuum limit) from a symmetric to a broken (or condensed) phase (which would be consistent with the  ``geometrogenesis'' scenario, if we had a full geometric interpretation for the simple TGFT model we are considering).  

\section{The model}
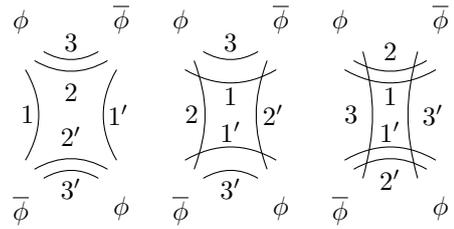
\begin{figure}
\centering
\begin{tikzpicture}[scale=0.6]
\path coordinate (1rs1) at (0,1)
      coordinate (1cs1) at (0,-1)
      coordinate (2rs1) at (0.2,1.2) 
      coordinate (2cs1) at (1.8,1.2) 
      coordinate (3rs1) at (0.4,1.4) 
      coordinate (3cs1) at (1.6,1.4) 
      coordinate (11rs1) at (2,-1) 
      coordinate (11cs1) at (2,1) 
      coordinate (21rs1) at (1.8,-1.2) 
      coordinate (21cs1) at (0.2,-1.2)
      coordinate (31rs1) at (1.6,-1.4) 
      coordinate (31cs1) at (0.4,-1.4)
      node (11) at (0.4,0) [left] {$1$}
      node (111) at (1.6,0) [right] {$1'$}
      node (31) at (1,1.2) [above] {$3$}
      node (21) at (1,0.9) [below]{$2$}
      node (211) at (1,-0.9) [above]{$2'$}
      node (311) at (1,-1.2) [below] {$3'$}
      (-0.1,1.3) node (phi11) [shape=circle,  label=above:$\phi$]{}
      (-0.1,-1.3) node (ovphi11) [shape=circle,  label=below:$\ov{\phi}$]{}
      (2.1,1.3) node (ovphi21) [shape=circle,  label=above:$\ov{\phi}$]{}
      (2.1,-1.3) node (phi21) [shape=circle,  label=below:$\phi$]{};
      \draw (1rs1) to [bend left=30]  (1cs1);
      \draw (2rs1) to [bend right=30]  (2cs1);
      \draw (3rs1) to [bend right=30]  (3cs1);
      \draw (11rs1) to [bend left=30] (11cs1);
      \draw (21rs1) to [bend right=30] (21cs1);
      \draw (31rs1) to [bend right=30] (31cs1);

\path[shift={(3.5,0)}] coordinate (1rs2) at (0,1) 
      coordinate (11cs2) at (0,-1)
      coordinate (2rs2) at (0.2,1.2) 
      coordinate (21cs2) at (1.8,1.2) 
      coordinate (3rs2) at (0.4,1.4)
      coordinate (3cs2) at (1.6,1.4) 
      coordinate (11rs2) at (2,-1) 
      coordinate (1cs2) at (2,1) 
      coordinate (21rs2) at (1.8,-1.2) 
      coordinate (2cs2) at (0.2,-1.2) 
      coordinate (31rs2) at (1.6,-1.4) 
      coordinate (31cs2) at (0.4,-1.4) 
      node (22) at (0.5,0) [left] {$2$}
      node (212) at (1.5,0) [right] {$2'$}
      node (32) at (1,1.2) [above] {$3$}
      node (12) at (1,0.8) [below]{$1$}
      node (112) at (1,-0.8) [above]{$1'$}
      node (312) at (1,-1.2) [below] {$3'$}      
      (-0.1,1.3) node (phi12) [shape=circle,  label=above:$\phi$]{}
      (-0.1,-1.3) node (ovphi12) [shape=circle,  label=below:$\ov{\phi}$]{}
      (2.1,1.3) node (ovphi22) [shape=circle,  label=above:$\ov{\phi}$]{}
      (2.1,-1.3) node (phi22) [shape=circle,  label=below:$\phi$]{};
      \draw (1rs2) to [bend right=30]  (1cs2);
      \draw (2rs2.south east) to [bend left=20]  (2cs2.north east);
      \draw (3rs2) to [bend right=30]  (3cs2);
      \draw (11rs2) to [bend right=30] (11cs2);
      \draw (21rs2.north west) to [bend left=20] (21cs2.south west);
      \draw (31rs2) to [bend right=30] (31cs2);
      
\path[shift={(7,0)}] coordinate (1rs3) at (0,1) 
      coordinate (11cs3) at (0,-1) 
      coordinate (2rs3) at (0.2,1.2) 
      coordinate (2cs3) at (1.8,1.2) 
      coordinate (3rs3) at (0.4,1.4) 
      coordinate (31cs3) at (1.6,1.4) 
      coordinate (11rs3) at (2,-1) 
      coordinate (1cs3) at (2,1) 
      coordinate (21rs3) at (1.8,-1.2) 
      coordinate (21cs3) at (0.2,-1.2) 
      coordinate (31rs3) at (1.6,-1.4) 
      coordinate (3cs3) at (0.4,-1.4) 
      node (33) at (0.5,0) [left] {$3$}
      node (313) at (1.5,0) [right] {$3'$}
      node (23) at (1,1) [above] {$2$}
      node (13) at (1,0.8) [below]{$1$}
      node (113) at (1,-0.8) [above]{$1'$}
      node (213) at (1,-1) [below] {$2'$}
      (-0.1,1.3) node (phi13) [shape=circle,  label=above:$\phi$]{}
      (-0.1,-1.3) node (ovphi13) [shape=circle,  label=below:$\ov{\phi}$]{}
      (2.1,1.3) node (ovphi23) [shape=circle,  label=above:$\ov{\phi}$]{}
      (2.1,-1.3) node (phi23) [shape=circle,  label=below:$\phi$]{};
      \draw (1rs3) to [bend right=30]  (1cs3);
      \draw (2rs3) to [bend right=30]  (2cs3);
      \draw (3rs3) to [bend left=15]  (3cs3);
      \draw (11rs3) to [bend right=30] (11cs3);
      \draw (21rs3) to [bend right=30] (21cs3);
      \draw (31rs3) to [bend left=15] (31cs3);
\end{tikzpicture}
\caption{Colored symmetric interaction terms.}
\label{Sym}
\end{figure}

We consider a rank-3 TGFT defined over 
 $\mathbb{R}^{ 3}$ endowed with a specific $\phi^4$ interaction called ``melonic'' \cite{bgrr}, shown in Fig.\ref{Sym}. In general, rank-3 melonic interactions correspond to peculiar triangulations of the 3-sphere
and are the most dominant objects in the large cut-off $N$ limit \cite{bgr,gurauryan,bgr2} (in both simple tensor models and topological GFTs, but this result is expected to extend to a wider class of models).
Written in momentum space\footnote{We adopt the standard QFT terminology for field modes, even though no spacetime interpretation is associated to the domain of the fields, and thus no standard physical interpretation should be associated to their modes. The same remark applies to our use of the terms `UV' and `IR' throughout the article.}, the classical action of the model reads:
\begin{eqnarray}
\label{model}
&&S[\phi,\ov{\phi}]=\int_{\mathbb{R}^{ 3}}\upd \textbf{p}\,
    \ov{\phi}_{123}\biggl(\sum_s p_s^2 +\mu\biggr)\phi_{123}\\
&&\hspace{-0.6cm}+\frac{\lambda}{2}\int_{\mathbb{R}^{ 6}}
\upd \textbf{p} \upd \textbf{p}' \, \biggl[
    \phi_{123}\ov{\phi}_{1'23}\phi_{1'2'3'}\ov{\phi}_{12'3'}
+\Sym\biggr],
\nonumber
\end{eqnarray}
where we used the notation $\phi_{123}=\phi(p_1,p_2,p_3)$ for the field modes and ``$\text{sym}$'' indicates that we include all the interactions obtained by symmetrisation over the color labels (see Fig.\ref{Sym}). The kinetic term is defined by a sum of Laplacians acting on the field indices and a mass term with coupling $\mu$. It is immediate to see that the action is built using generalised traces over field indices convoluted with the kinetic and interaction kernels and, once exponentiated, defines a quantum theory through a Gaussian field measure of covariance $(\sum_s p_s^2 +\mu)^{-1}$.

\section{FRG equations for tensorial models}

The Functional Renormalisation Group approach \cite{delamotte,wetterich,wetterich2,morris} rephrases the problem of integrating out the high modes of a theory, as one of solving a differential equation, the FRG equation. Being non-perturbative in nature 
(with respect to any expansion in the interaction coupling constants), the FRG allows in principle to deal with the full set of quantum fluctuations of the model and to study its critical behavior.

The implementation of the FRG method in TGFT \cite{dariojosephdaniele} follows closely the usual one \cite{delamotte,wetterich,morris}, with special attention payed to the fact that we are dealing with a convolution of tensors and thus with peculiar non-local interactions.
We start by decoupling the field modes as typical in the Wilsonian approach to renormalisation, by adding to the action a mass-like regulator term 
$\Delta S_k=\Tr(\ov{\phi}\cdot R_k\cdot\phi)$
depending on the IR cut-off $k$, that splits the modes into high modes ($|\textbf{p}|>k$) and low modes ($|\textbf{p}|<k$). The scale dependent quantum theory will then be defined through a partition function where high modes are integrated out:
\begin{equation}
Z_k[J,\ov{J}]=\int\upd\phi\upd\ov{\phi}\;e^{-S[\phi,\ov{\phi}]-\Delta S_k[\phi,\ov{\phi}]
   +\Tr(\ov{J}\phi)+\Tr(J\ov{\phi})}\,,
\end{equation}
where $J$ is a complex tensor playing the role of a source and
$\Tr(J \ov{\phi}) := \int _{\mathbb{R}^{ 3} } J_{123} \ov{\phi}_{123}$. 
After a Legendre transform, we identify a scale dependent effective action which encodes the full information about the quantum theory: 
\begin{equation}
\Gamma_k[\varphi,\ov{\varphi}]=
\sup_{J,\ov{J}}\{\Tr(J\ov{\varphi})+\Tr(\ov{J}\varphi)-W_k[J,\ov{J}]
   -\Delta S_k[\varphi,\ov{\varphi}]\},
\end{equation}
where $\varphi=\langle\phi\rangle$ and $W_k[J,\ov{J}]=\log Z_k[J,\ov{J}]$. The  term $\Delta S_k$ 
is also chosen to be compatible with the choice of initial conditions for the FRG differential equation, encoding the scaling of effective action to the bare one in the UV:
$\Gamma_k[\varphi,\ov{\varphi}]\underset{k\to\Lambda}{\longrightarrow}S[\varphi,\ov{\varphi}]$, where $\Lambda$ plays the role of a UV cut-off.
Introducing the logarithmic scale $t=\log k$ and $ \Gamma_k^{(2)}:=\delta\Gamma_k/\delta\ov{\varphi}\delta\varphi$, the Wetterich equation for tensorial GFT models has the form \cite{dariojosephdaniele}:
\begin{equation}
\label{Wett}
\de_t\Gamma_k=\Tr(\de_tR_k\cdot[\Gamma_k^{(2)}+R_k]^{-1})\,. 
\end{equation}
Equation \eqref{Wett} is fully non-perturbative and exact, and encodes (formally) all the information about quantum fluctuation in a typical one-loop form. 

Before moving on to the solution of this equation and to the study of the critical points, let us anticipate an important technical fact, which we will have to deal with in the following. Despite the presence of momentum cut-offs, evaluating \eqref{Wett} requires, as in the ordinary scalar field theory, an infinite volume regularisation. In the local field theory case, infinite volume divergences are cured by passing to constants field modes or taking a thermodynamic limit \cite{wetterich,delamotte}. In the case of compact groups as worked out in \cite{dariojosephdaniele}, the issue does not arise, as the group volume is finite. The presence of a finite radius results in a non-autonomous system of equations where the scale $k$ appears explicitly. The existence of phase transition can be inferred only in the limit of infinite radius \cite{dariocurved}. The present situation differs from both cases, namely the local field theory and compact TGFTs in the limit of infinite volume. Because of the crucial non-local properties of the interactions, the use of constant field modes is misleading (indeed $\phi^4$ terms in the TGFT case do not have the same combinatorics, and this cannot be neglected) and, in the compact group case, one must understand how to best perform, both from a conceptual and practical point of view, an infinite radius limit in the equations. In our case, we then find wiser to perform a thermodynamic limit. 

\section{Truncation scheme}

In order to be able to perform practical computations, we need to adopt a truncation scheme for the effective action. 

Of course performing a truncation means loosing the exact nature of the Wetterich equation. Generally, this also generates a singularity of the flow that splits the space of couplings in disconnected regions. In a neighbourhood of the singularity, we cannot trust the computations. Since one usually is interested in the free theory around which the perturbative expansion makes sense, we will discuss only the region connected to the origin in the space of couplings. 

We choose to truncate $\Gamma_k$ to the quadratic term in the derivative of the fields and to order four in the fields, thus obtaining a form similar to the action itself:
\begin{eqnarray}
\label{GAMMA}
&&\Gamma_k [\varphi,\ov{\varphi}]=\int_{\mathbb{R}^{ 3}} \upd \textbf{p}\;
    \ov{\varphi}_{123}\biggl(Z_k\sum_s p_s^2 +\mu_k\biggr)\varphi_{123}\\
&&\hspace{-0.6cm}+\frac{\lambda_k}{2}\int_{\mathbb{R}^{ 6}}
\upd \textbf{p} \upd \textbf{p}' \, \;\biggl[
    \varphi_{123}\ov{\varphi}_{1'23}\varphi_{1'2'3'}\ov{\varphi}_{12'3'}
+\Sym\biggr].
\nonumber
\end{eqnarray}
We can already see that, in this way, the UV initial condition on the flow is satisfied.

From \eqref{GAMMA}, the 2-point 1PI Green function expresses as
$\Gamma_k^{(2)}(\textbf{p},\textbf{p}')=(Z_k\sum_sp_s^2+\mu_k)\delta(\textbf{p}-\textbf{p'})+F_k(\textbf{p},\textbf{p}')$, where
\begin{align}\label{fterm}
&F_k(\textbf{p},\textbf{p}')=\lambda_k\biggl[\int dq_1\
     \varphi_{q_1p'_2p'_3}\ov{\varphi}_{q_1p_2p_3}\delta(p_1-p'_1)\crcr
&+\int dq_2dq_3\;\varphi_{p'_1q_2q_3}\ov{\varphi}_{p_1q_2q_3}
   \delta(p_2-p_2')\delta(p_3-p'_3)\crcr
&+\Sym\biggr]\,. 
\end{align}
The regulator function is chosen as \cite{litim2}:
$R_k(\textbf{p},\textbf{p'})=\delta(\textbf{p}-\textbf{p'})Z_k(k^2-\sum_{s=1}^3p_s^2)
   \theta(k^2-\sum_{s=1}^3p_s^2)$,
where $\theta$ stands for the Heaviside step function. This is a standard choice and it satisfies all basic requirements, namely:
$R_{k=0}=0,\;\forall\,\textbf{p}$, so that
$Z_{k=0}[J,\ov{J}]=Z[J,\ov{J}]$; $R_{k=\Lambda}\varpropto\Lambda^2,\;\forall\,\textbf{p}\;\text{s.t.}\;|\textbf{p}|<k$,
to approximately freeze the propagation of modes with norm smaller than $k$;
$R_k(|\textbf{p}|>k)=0$, so that high modes are unaffected by the regulator. In addition, this choice is particularly interesting in
our framework because its functional properties allow the analytic evaluation of spectral sums. 

If we act on the regulator with the derivative with respect to the logarithmic scale, we find 
$\de_tR_k=\theta(k^2-\Sigma_sp_s^2)[\de_tZ_k(k^2-\Sigma_sp_s^2)+2k^2Z_k]\delta(\textbf{p}-\textbf{p}')$
and the $\delta(k^2 - \Sigma_sp_s^2)$-term so generated simply cancels out.

Expanding the Wetterich equation, it seems natural to choose an expansion in powers of $(\varphi\ov{\varphi})$, which we perform up to the third order, and discarding the vacuum terms, to obtain our final truncated functional equation, from which we read out the differential equations for the beta functions of the theory. 

\section{Thermodynamic limit}

In order to regularise volume divergences, we perform a lattice regularisation
in the $\textbf{p}$-space, which follows from a compactification in the direct space, according to the conventions of \cite{salm}. 
Defining the model \eqref{model} over a lattice 
$D^*=[\frac{2\pi}{L}\mathbb{Z}]^{ 3}= [\frac{1}{r}\mathbb{Z}]^{ 3}
:=[l\mathbb{Z}]^{ 3}$,
of spacing $l^3$ proportional to the volume of the direct space, the Fourier transform becomes a Fourier series
and, for any function $f(\textbf{p})$, we have
$
 \int_{D^*}dp_i\;f(\textbf{p})=l^3\sum_{\{p_i\}\in D^*}f(\textbf{p})
$.
We define the delta distribution in ${D^*}$ as:
$ \delta_{D^*}(\textbf{p},\textbf{q})=\delta_{\textbf{p},\textbf{q}}/l^3$\,,
with $\delta_{\textbf{p},\textbf{q}}$, the Kronecker delta. As a result, we have:
  $\delta_{D^*}(\textbf{p},\textbf{p})=\delta_{\textbf{p},\textbf{p}}/l^3=1/l^3$.

Using this regularisation prescription, the effective action of the model 
reads: 
\begin{eqnarray}
&& \Gamma_k[\varphi,\ov{\varphi};l]=
   l^3\sum_{\textbf{p}\in D^*}\ov{\varphi}_{123}\Big(Z_k\sum_sp_s^2+\mu_k\Big)\varphi_{123}
\\
&&\hspace{-0.5cm}+\frac{l^6\lambda_k}{2}\sum_{\textbf{p},\textbf{p}'\in D^*}\biggl[
    \varphi_{123}\ov{\varphi}_{1'23}\varphi_{1'2'3'}\ov{\varphi}_{12'3'}
+\Sym\biggr],
\nonumber
\end{eqnarray}
where   $\varphi(\textbf{p})=\int_D[dx_i]_{i=1}^3\;e^{-i\sum_ip_ix_i}\varphi(x_1,x_2,x_3)$. 
In the end, the continuous description will be recovered in the 
thermodynamic limit $l\to 0$. 

The dependence of the system on the volume of the direct space is now explicit, and we can tune this dependence in order to consistently remove all the divergences, and be left with he physical $\beta$-functions.

\section{$\beta$-functions}

The IR regularisation of the system of $\beta$-functions is direct: we need to extract from the coupling constants
an explicit dependence on the volume of the direct space, in addition to their scaring with the momentum cut-off. 
After a lengthy but straightforward calculation, the set of $\beta$-functions computed with the prescription introduced in the previous section reads\footnote{We drop here the symbol $\textrm{lim}_{l\to 0}$ to simplify the notation.}:
\begin{eqnarray}
\beta(Z_k)&=&\frac{\lambda_k}{(Z_kk^2+\mu_k)^2}
     \biggl[2Z_k\biggl(2\frac{k}{l}+\pi \frac{k^2}{l^2}\biggr)\crcr
&&+\de_t Z_k\biggl(\pi \frac{k^2}{l^2}+4\frac{k}{l}\biggr)\biggr]\,,\crcr
\beta(\mu_k)&=&-3\frac{\lambda_k}{(Z_kk^2+\mu_k)^2}
     \biggl[2Z_k\biggl(\frac{k^4}{l^2}\pi+2\frac{k^3}{l}\biggr)\crcr
&&+\de_t Z_k\biggl(\frac{\pi}{2}\frac{k^4}{l^2}+\frac{4}{3}\frac{k^3}{l}\biggr)\biggr]\,,\crcr
\beta(\lambda_k)&=&\frac{2\lambda_k^2}{(Z_kk^2+\mu_k)^3}\biggl[
     2Z_k\biggl(\pi \frac{k^4}{l^2}+10\frac{k}{l}+2k^2\biggr)\crcr
&&+\de_t Z_k\biggl(\frac{\pi}{2}\frac{k^4}{l^2}+\frac{20}{3}\frac{k^3}{l}+2k^2\biggr)\biggr]\,.
\end{eqnarray}
To make sense of it in the infinite volume limit,
we use the ansatz:
\begin{equation}
Z_k=\ov{Z}_kl^{\chi}k^{-\chi},\quad\mu_k=\ov{\mu}_k\ov{Z}_kl^{\chi}k^{2-\chi},
   \quad\lambda_k=\ov{\lambda}_k\ov{Z}_k^2l^{\xi}k^{\sigma}\,,
\end{equation}
where $[\ov{Z}_k]=[\ov{\mu}_k]=[\ov{\lambda}_k]=0$, $[\varphi]=-\frac{5}{2}$ and $\xi+\sigma=4$. These dimensions are fixed by requiring that $[\Gamma_k]=0$ with $[p]=1$.
Now, extracting the dimensionless $\beta$-functions, one gets:
\begin{align}
\label{systembeta}
\eta_k&={\textstyle\frac{\ov{\lambda}_kl^{\xi}k^{\sigma}}{l^{2\chi}k^{4-2\chi}(1+\ov{\mu}_k)^2}}
   \biggl[(\eta_k-\chi)\biggl(4\frac{k}{l}+\pi\frac{k^2}{l^2}\biggr)\crcr
&+2\biggl(2\frac{k}{l}+\pi\frac{k^2}{l^2}\biggr)\biggr]+\chi\,,\crcr
\beta(\ov{\mu}_k)&=-{\textstyle\frac{3\ov{\lambda}_kl^{\xi}k^{\sigma}}{l^{3\chi}k^{6-3\chi}(1+\ov{\mu}_k)^2}}
   \biggl[(\eta_k-\chi)\biggl(\frac{\pi}{2}\frac{k^4}{l^2}+\frac{4}{3}\frac{k^3}{l}\biggl)\crcr
&+2\biggl(\pi\frac{k^4}{l^2}+2\frac{k^3}{l}\biggr)\biggr]-\eta_k\ov{\mu}_k-(2-\chi)\ov{\mu}_k\,,\crcr
\beta(\ov{\lambda}_k)&=
   {\textstyle\frac{2\ov{\lambda}_k^2l^{\xi}k^{\sigma}}{l^{3\chi}k^{6-3\chi}(1+\ov{\mu}_k)^3}}
   \biggl[(\eta_k-\chi)\biggl(\frac{\pi}{2}\frac{k^4}{l^2}+\frac{20}{3}\frac{k^3}{l}+2k^2\biggr)\crcr
&+2\biggl(\pi\frac{k^4}{l^2}+10\frac{k}{l}+2k^2\biggr)\biggr]-2\eta_k\ov{\lambda}_k-\sigma\ov{\lambda}_k\,.
\end{align}
The system \eqref{systembeta} of $\beta$-functions is non-autonomous in the IR cut-off $k$ as long as the parameter $l$ is kept finite. This feature is due to the peculiar combinatorics of the tensorial vertices which span the 1PI 2-point functions with different volume contributions. One way to realise this is by noting the unusual delta distributions in $F$ in \eqref{fterm}. From \eqref{systembeta}, 
we see two different systems arising in the UV and IR cut-off limits, coming from different leading terms. The fact that the set of $\beta$-functions of a TGFT over a compact group manifold is non-autonomous  is consistent with the analysis of standard field theories on compact (and curved) manifolds \cite{dariocurved}.

In order to make sense of the non-compact limit, we solve the system in the variables $\xi$ and $\chi$ by requiring that the highest volume contribution is regularised and all the sub-leading infinities are sent to zero. We have:
$
\Big\{\begin{aligned}
&\xi-2\chi-2=0\\
&\xi-3\chi-2=0
\end{aligned}
$ which yields $\chi=0$, $\xi=2$, $\sigma=2$.  
The resulting system of differential equations for the theory is:
\begin{equation}
\label{betaad}
\left\{
\begin{aligned}
\eta_k=&\frac{\pi\ov{\lambda}_k}{(1+\ov{\mu}_k)^2}(\eta_k+2)\\
\beta(\ov{\mu}_k)=&-\frac{3\pi\ov{\lambda}_k}{(1+\ov{\mu}_k)^2}(\frac{\eta_k}{2}+2)
     -\eta_k\ov{\mu}_k-2\ov{\mu}_k\\
\beta(\ov{\lambda}_k)=&\frac{\pi\ov{\lambda}_k^2}{(1+\ov{\mu}_k)^3}(\eta_k+4)
     -2\eta_k\ov{\lambda}_k-2\ov{\lambda}_k\\
\end{aligned}
\right.
\end{equation}
which is the starting point of our computation of the RG flow. As expected, absent any remaining fixed external scale, the system is now autonomous.

\section{The RG flow}

Proceeding with the standard analysis, we first determine the fixed points and then study the linearised system around them to determine the critical exponents of the model.
From the non-linear nature of the $\beta$-functions, we have a singularity at $\ov{\mu}_k=-1$ and $\ov{\lambda}_k=(1+\ov{\mu}_k)^2/\pi$.
In a neighbourhood of those singularities, we do not trust the linear approximation and, being interested mainly in the sector of the theory connected with the Gaussian fixed point (i.e. to the perturbative regime of the theory), we will not study the flow around points beyond the singularities. 
By numerical evaluation, we find a Gaussian fixed point (GFP) and three non-Gaussian (NGFP) fixed points in the plane $(\ov{\mu}_k,\ov{\lambda}_k)$. We discard one of them because it lies beyond the singularity. The others correspond to
\begin{equation}
 P_1=(8.619,-47.049),    \quad P_2=10^{-1}(-6.518,0.096) \;\; \nonumber. 
\end{equation}
The stability matrix at the GFP has an eigenvalue with algebraic multiplicity 2 corresponding to the canonical scaling dimensions of the couplings:
$\theta_{1,2}^G=-2$, but one single eigenvector $\textbf{v}=(1,0)$, thus, considering that all the trajectories flow into the origin,
the GFP must have a marginal direction in the UV.
In a neighborhood of the non-Gaussian fixed points, we have:
\begin{align}
 &\theta_+^1\sim0.351\text{ for }\textbf{v}_+^1\sim10^{-1}(0.65,-9.98),\\
   &\theta_-^1\sim-2.548\text{ for }\textbf{v}_-^1\sim10^{-1}(-6.88,7.26),\\
&\theta_+^2\sim10.066\text{ for }\textbf{v}_+^2\sim10^{-1}(9.996,-0.269),\\
   &\theta_-^2\sim-1.988\text{ for }\textbf{v}_-^2\sim10^{-1}(9.987,0.506).
\end{align}
\begin{figure}
 \centering
\includegraphics[scale=0.28]{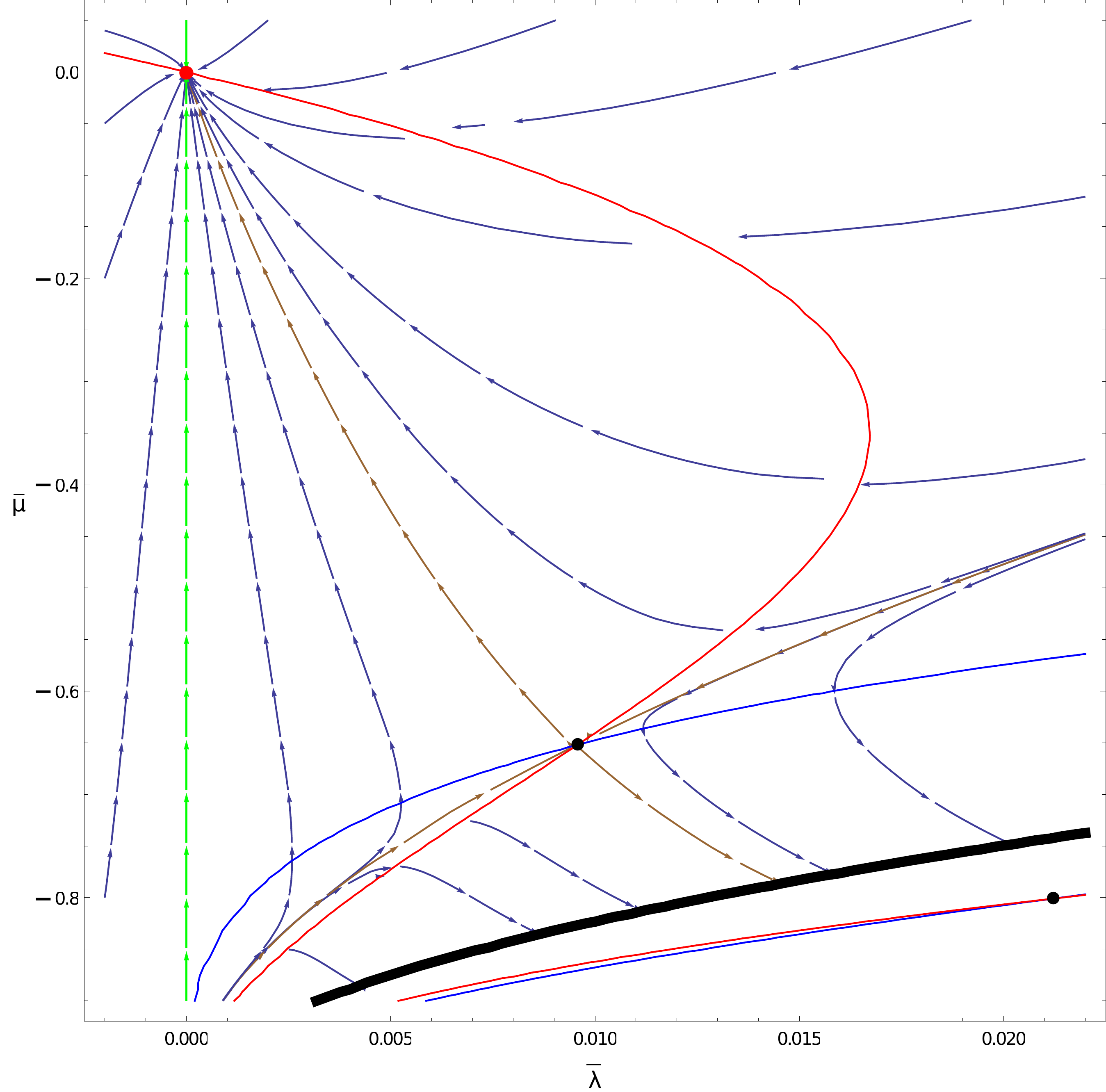}
\includegraphics[scale=0.26]{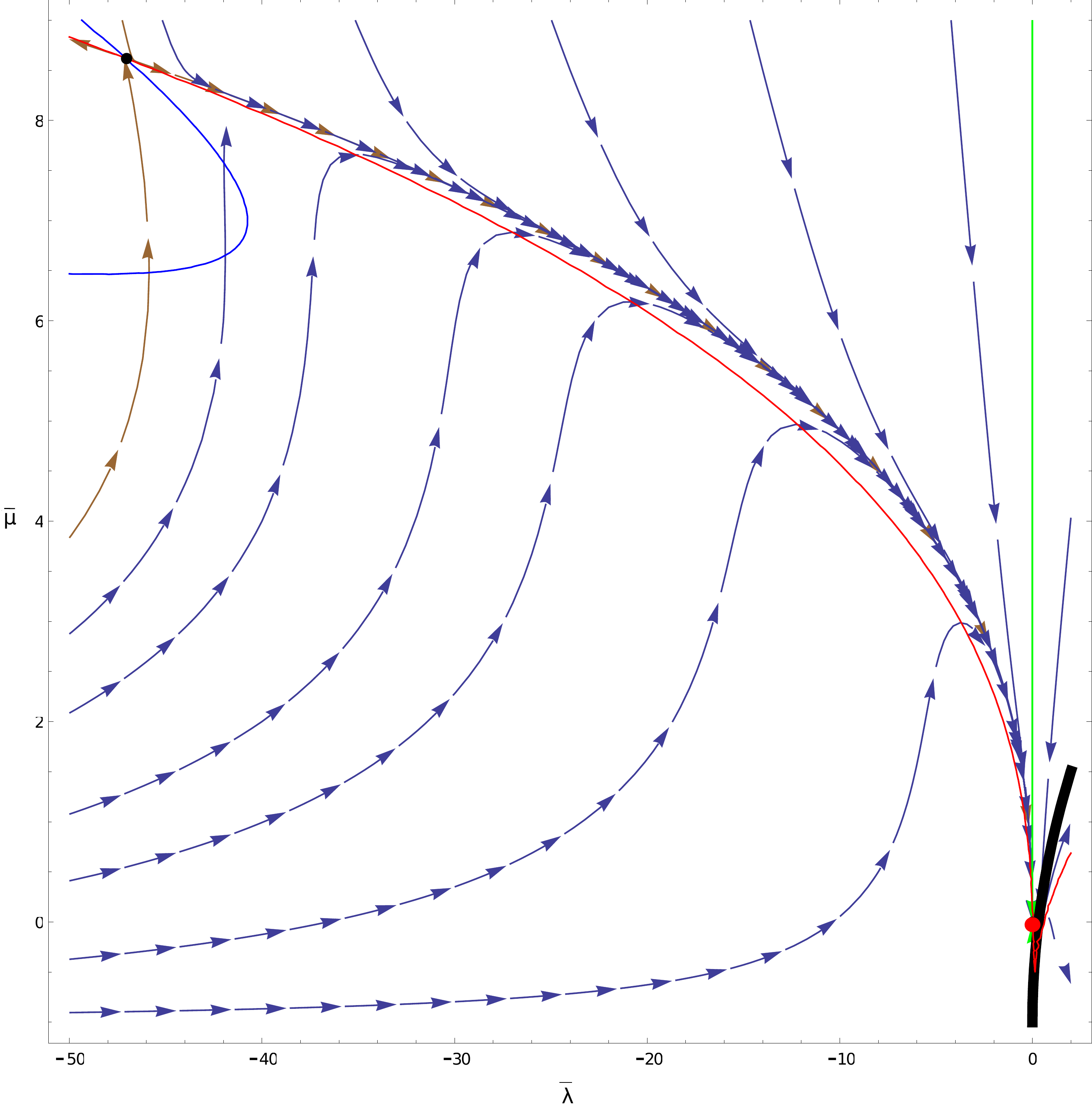}
\caption{RG flow of the model - The red and blue lines represent, respectively, the zeros of $\beta(\ov{\mu}_k)$ and $\beta(\ov{\lambda}_k)$, the brown arrows are the eigenperturbations of the non-Gaussian fixed points (in black), while the green ones those of the Gaussian fixed point (in red). Arrows point in the UV direction. The thick black line is the singularity of the flow.}
\label{plots}
\end{figure}
The flow of the couplings between the two NGFPs and the Gaussian one are plotted in Fig.\ref{plots}.  The origin is a UV sink for the flow; hence, the model is asymptotically free. As mentioned before, 
the absence of a second eigenvector for the stability matrix around the GFP requires an approximation beyond the linear order and is a signal of the presence of a marginal perturbation. 
By close inspection of the plots, confirmed by direct 
 integration at second order of the system of $\beta$-functions, which can be performed for generic numerical constants/initial conditions, we infer that the behaviour of this direction is still UV attractive, i.e. that it corresponds to a marginally relevant direction.

Both the non-Gaussian fixed points have one relevant and one irrelevant direction. 
They are also characterised by the so-called ``large river effect''. 
This effect shows a splitting of the space of coupling in two regions not connected by any RG 
trajectory. Thus, the irrelevant direction for the NGFP match the properties of a critical surface and suggests
the presence of phase transitions in the model.
In the $\ov{\lambda}>0$ plane, the flow is similar to the one of standard 
local scalar field theory on $\mathbb{R}^3$ in a neighbourhood of the
Wilson-Fisher fixed point. That is: above the critical surface, the IR limit of the RG trajectories brings the theory in a region where both $\ov{\mu}_k$ and $\ov{\lambda}_k$ are positive, while below the irrelevant eigendirection for $P_2$, the mass parameter is driven to be negative in the IR, indicating a spontaneous symmetry breaking mechanism (in the different but related context of tensor models, such a mechanism has been also found in \cite{Benedetti:2015ara}). In the sector $\ov{\lambda}<0$, the situation
is rather peculiar. We might infer that $P_1$ has the same properties just discussed but reversed with respect to the critical surface. 
The symmetric phase where $\ov{\mu}_k$ and $\ov{\lambda}_k$ have the same sign in the IR is below the irrelevant direction of the fixed point, while the broken phase lies above it. In this sense, we have a phase transition also crossing the surface $\ov{\lambda}=0$, but this is not an irrelevant direction for any NGFP. This feature suggests that, in this case, we may have a first order phase transition. Nevertheless, we must remember that the sector $\ov{\lambda}<0$ generates theories with a non-stable coupling, which is generally not considered in a field theory context. This sector must therefore be analyzed under a different parametrisation, if we want to shed more light on it.

\

In a GFT model with additional geometric data, and a proper simplicial gravity interpretation, a broken or condensate phase could be interpreted as a continuum geometric phase \cite{danieleEmergence,GFTcondensate}, and would support a geometrogenesis scenario for the emergence of continuum spacetime and geometry from these GFT models.  
The model under consideration would therefore need to be enriched with such additional data to be more than an indirect support
for such a scenario. 
Also in our model, in any case, a proper study of the broken phase, involving a change in parametrisation for the effective potential and a detailed study
of the theory around the new ground state, solving the classical equation of motion of the model, in a saddle point approximation, would be needed to confirm conclusively the existence of a phase transition as envisaged. 

\acknowledgments
The authors are thankful to the Albert Einstein Institute and the University of Bologna for having made possible this collaboration. They are also very much grateful to Dario Benedetti and Astrid Eichhorn for helpful discussions and several useful comments. R.M. warmly thanks the AEI for
its  hospitality.

\

\end{document}